\documentstyle[11pt,epsfig,amssymb,citesort]{article}
\newcommand{\beq}{\begin{eqnarray}}
\newcommand{\eeq}{\end{eqnarray}}

\newcommand{\np}{Nucl. Phys.\ }
\newcommand{\pl}{Phys. Lett.\ }

\newcommand{\asgen}{\alpha_s}

\newcommand{\astil}{\alpha_s{{}^{\widetilde{\rm MOM}}}}

\newcommand{\MOM}{\widetilde{\rm MOM}}

\newcommand{\Lams}{\Lambda_{\overline{\rm MS}}}

\newcommand{\be}{\begin{equation}}
\newcommand{\ee}{\end{equation}}
\newcommand{\lwrsim}{\raise0.3ex\hbox{$<$\kern-0.75em\raise-1.1ex\hbox{$\sim$}}}
\def\Am#1#2#3{\widetilde A_{#1}^{#2}(#3)}
\def\A#1#2#3{A_{#1}^{#2}(#3)}
\def\C2#1#2{({\cal C}_2)_{#1}^{#2}}

\def\jhep#1#2#3{J. High Energy Phys. {\bf #1} (#2) #3}
\def\prd#1#2#3{Phys.\ Rev.\ {\bf D#1} (#2) #3}
\def\npb#1#2#3{Nucl.\ Phys.\ {\bf B#1} (#2) #3}
\def\plb#1#2#3{Phys.\ Lett.\ {\bf B#1} (#2) #3}

\def\np#1#2#3{Nucl.\ Phys.\ B#1 (19#3) #2}
\def\pl#1#2#3{Phys.\ Lett.\ #1B (19#3) #2}

\def\zpc#1#2#3{Z.\ Phys.\ {\bf C#1} (#2) #3}
\begin{document}
%\today
\setcounter{page}{1}
\begin{flushright}
LPT-ORSAY 01/05\\
UHU-FT/01-01\\
\end{flushright}
%filename="puisalpha2.tex"
\begin{center}
\bf{\huge 
Testing Landau gauge OPE on the Lattice with a $\langle A^2\rangle$ Condensate}
\end{center}  
\vskip 0.8cm
\begin{center}{\bf  Ph. Boucaud$^a$, A. Le Yaouanc$^a$, J.P. Leroy$^a$, 
J. Micheli$^a$, \\ 
O. P\`ene$^a$, J. Rodr\'{\i}guez--Quintero$^b$   
}\\
\vskip 0.5cm 
$^{a}$ {\sl Laboratoire de Physique Th\'eorique~\footnote{Unit\'e Mixte 
de Recherche du CNRS - UMR 8627}\\
Universit\'e de Paris XI, B\^atiment 210, 91405 Orsay Cedex,
France}\\
$^b${\sl Dpto. de F\'{\i}sica Aplicada e Ingenier\'{\i}a el\'ectrica \\
E.P.S. La R\'abida, Universidad de Huelva, 21819 Palos de la fra., Spain} \\
\end{center}
\begin{abstract}
\medskip

Using the operator product expansion we show that the  $O(1/p^2)$ correction to
the perturbative expressions for the gluon propagator and the strong coupling
constant resulting from lattice simulations  in the Landau gauge are due to a
non-vanishing vacuum expectation  value of the  operator $A^\mu A_\mu$. This is
done using the  recently published Wilson coefficients of the identity operator
computed to third order, and the subdominant  Wilson coefficient computed in this
paper to the leading logarithm. As a test of the applicability of OPE we compare
the  $\langle A^\mu A_\mu\rangle$ estimated from the gluon propagator and the one from the
coupling constant in the flavourless case. Both agree within the statistical uncertainty: $\sqrt{\langle A^\mu
A_\mu \rangle} \simeq 1.64(15)$ GeV. Simultaneously we fit $\Lams = $ 233(28) MeV  in perfect
agreement with previous lattice estimates. When the leading coefficients are 
only expanded to two loops, the two estimates of the condensate differ
drastically. As a consequence we insist that OPE can be applied in predicting
physical quantities only if the Wilson coefficients are computed to a high enough
perturbative order.

\noindent P.A.C.S.: 12.38.Aw; 12.38.Gc; 12.38.Cy; 11.15.H

\end{abstract}

\section{Introduction}
\label{sec:intro}

\hspace*{\parindent} When computing in a fixed gauge an operator product,  the
operator product expansion (OPE) contains in general contributions from local
gauge-dependent operators, even though  they should not emerge in the
gauge-invariant sector.  For example in ref. \cite{lavelle}, a detailed analysis 
clearly shows that operators such as  $A^2=A_\mu A^\mu$  contribute to QCD
propagators' OPE through a non-zero expectation value in a non-gauge-invariant
``vacuum''. $A^2$ is the unique dimension two operator allowed to have a vacuum
expectation value (v.e.v.) and is thus  the dominant non perturbative
contributor, leading to $\sim 1/p^2$ corrections to the perturbative result.  

These expected $\sim 1/p^2$ have at first sight nothing to do  with the possible
presence of $1/p^2$ terms in {\it gauge invariant} quantities such as Wilson
loops \cite{Ceccobeppe}:  since no  local gauge invariant gluonic operator of
dimension less than 4 exists  it
is expected from OPE that the dominant power correction should be $\propto
1/p^4$, originating from the local and gauge-invariant $G^{\mu \nu} G_{\mu \nu}$. 
Of course  the operator $A^2$ in the Landau gauge can be viewed, by
simply averaging it over the gauge orbit, as a gauge invariant non local
operator.  But then, dealing with non-local operators, we loose the standard OPE
power counting rule relating the power behaviour  of a Wilson coefficient to the
dimension of the corresponding operator:  there is no reason for this non-local
operator  to yield $1/p^2$ contributions in a gauge invariant observable. 
 It has been strongly stressed  in ref. \cite{Zakh} that, working in the Landau
gauge, the $A^2$ operator  plays a special role since, imposing the Landau gauge
condition is equivalent to asserting that $A^2$ is at an extremum or a saddle
point on its gauge orbit.  Practically, on a lattice, one fixes the Landau  gauge
by searching for a minimum of $A^2$ on the orbit. We are not able to elaborate
further on the issue  of what relation might exist between the {\it expected}
$\langle A^2 \rangle$ condensate in the landau gauge and the possible {\it unexpected} $1/p^2$ terms in
gauge invariant quantities \cite{Ceccobeppe}. But we are in a position to put the
first step of this possible route on a firm ground: to provide a strong evidence
that  there is indeed an $\langle A^2\rangle$ condensate in the Landau gauge and that it is
not small. 

To that aim we will use heavily OPE.  We need to be sure that OPE really works in
this situation and have to invent some way of verifying this point.
  A success of this check would
achieve several goals. First it would  give a strong support to the conjecture 
that OPE is really working in this situation, i.e. that we do not encounter a
strange situation where OPE would have failed like the one discussed in the preceding
paragraph about $\langle G^{\mu \nu} G_{\mu \nu}\rangle$. Second it would confirm that we go
far enough in the perturbative expansion (the expansion in $1/\ln (p^2)$) to be
able to say something sensible about  the power expansion (in $1/p^2$). Third it
would confirm that we really  are measuring $\langle A^2\rangle$.  Such
checks have of  course many consequences which will be further discussed in the
conclusions. 

 From a 
 practical (numerical) point of view, $1/p^2$ terms provide a specially 
convenient way to test  OPE
 since 
\begin{itemize}  %\item[-] our lattice results have a high accuracy
 %\item[-] we can rely on recent progress in perturbative calculations 
%\cite{Chety2}
 \item[-]they  remain visible at much larger energies than the $1/p^4$ ones 
which would result from the gauge invariant $G^{\mu \nu} G_{\mu \nu}$ %and are the
%natural candidates  for the sizeable corrections we have found at energies $\sim
%10$ GeV. 
\item[-] as already mentioned their OPE analysis is rather simple and
unambiguous because $A^2$ is the only  dimension-two operator to  contribute. 
%$\asgen$  in the symmetric MOM scheme.
 \end{itemize} 

A recent study of  $\astil(p)$, the Landau gauge coupling constant\footnote{$\astil(p)$ stands for the QCD running coupling constant
non-perturbatively renormalized in a kinematically asymmetric point by following
the {\it momentum subtraction} prescriptions ($\MOM$)}, regularized on a lattice showed
 unequivocally the presence of $1/p^2$ power corrections still visible at
energies $\sim 10$ GeV for which OPE contributions of the gluon condensate,
$\langle A^2 \rangle$,  were natural
candidates\cite{poweral}. In this term all the non-perturbative input is contained in 
$\langle A^2\rangle$ while the OPE Wilson coefficients can be computed in perturbation. In view of  this, we
proposed in a previous work \cite{OPE} a  procedure to test OPE based on the
determination, and further comparison, of the two estimates of the gluon
condensate, $\langle A^2 \rangle$, obtained from both gluon two- and three-point
Green functions by means of a simultaneous matching of the lattice data  to the
OPE formulas derived by following standard Shifman-Vainshtein-Zakharov (SVZ)
techniques \cite{weinberg}. Thus,
our OPE matchings of lattice data provides two independent estimates of the
renormalized $A^2$ condensate. The adequate definition of renormalized
condensates and their ``universality'' when studying different Green functions was
discussed in ref. \cite{OPE} in connection with the choice of truncation orders
for perturbative and OPE series (see also \cite{david,MartiSach}). In this
preliminary work we described the theoretical framework for this testing
procedure and we performed a first analysis of previous lattice data
\cite{frenchalpha,propag}  but the perturbative $\beta$-function was known at that time only up to two loops and our use of OPE  was limited to the
sole computation of the  Wilson coefficients of $A^2$ at  {\it tree-level}

After this work was completed a computation of the  third coefficient of the MOM
beta function, $\beta_2$ has been published in  ref. \cite{Chety2}. The authors
of this last work conclude that their computed $\beta_2$  and our ``prediction"
of this coefficient based on OPE consistency~\cite{OPE} reasonably  agree with
each other.  Thanks to the new information concerning the $\beta$-function and to
the high accuracy of our lattice
 results we are now in a particularly favourable situation to address further the 
questions we have mentionned above. This is the task we shall attack in the
 present paper,  presenting a  consistent calculation  in the MOM
scheme (a symmetric kinematics chosen for the vertex) with the
Wilson coefficients of the identity operator computed at three loops
\cite{Davy,Chety2,frenchalpha,propag} and the ones of $A^2$  computed to the
leading logarithm in section~\ref{wilson}. In particular
we will compare the check of the  ``universality'' of the condensates when
expanding the leading perturbative coefficients to  three loops and when one
uses  only the two loop order.

 The theoretical setting of our use of OPE is described in Section 
\protect\ref{sec:model}: the tree-level computation, presented previously in ref.
\cite{OPE}, is only sketched and  most attention is paid to the obtention of the
one-loop anomalous  dimension of the Wilson coefficients. The fitting strategy is
explained and the matching test  performed in  \protect\ref{sec:lattice}.
Finally, we discuss and conclude in Section~\protect\ref{sec:conclusions}.

\section{OPE for the gluon propagator and  $\asgen(p)$}
\protect\label{sec:model}

\hspace*{\parindent}  In the present section we shall expand the three-point Green function,
and  hence $\alpha_S(p)$, as well as the gluon propagator,  in the   OPE
approach up to the the $1/p^2$ order. Both gluonic two- and three-point Green
functions are renormalized  according to the MOM scheme.  Let us start with  a
reminder of the computation of the tree-level Wilson coefficients~\cite{OPE}.

\subsection{Tree-level Wilson coefficients}

\hspace*{\parindent}  In the pure Yangs-Mills QCD, without quarks, OPE yields

\beq
T\left( \Am{\mu}{a}{-p} \Am{\nu}{b}{p}\right)&=& \nonumber \\
(c_0)^{a b}_{\mu \nu}(p) \ 1 
&+& (c_1)^{a b \mu'}_{\mu \nu a'}(p) : \A{\mu'}{a'}{0}: \ + \
(c_2)^{a b \mu' \nu'}_{\mu \nu a' b'}(p) \ 
:\A{\mu'}{a'}{0} \ \A{\nu'}{b'}{0}:  \nonumber \\
&+&\dots \;\; , \label{OPEfield1} 
\eeq
\beq
T\left( \Am{\mu}{a}{p_1} \Am{\nu}{b}{p_2} \Am{\rho}{c}{p_3} \right)&=& \nonumber \\
(d_0)^{a b c}_{\mu \nu \rho}(p_1,p_2,p_3) \ 1 
&+& (d_1)^{a b c \mu'}_{\mu \nu \rho a'}(p_1,p_2,p_3) : \A{\mu'}{a'}{0}:
\nonumber \\ 
&+& (d_2)^{a b c \mu' \nu'}_{\mu \nu \rho a' b'}(p_1,p_2,p_3) \   
:\A{\mu'}{a'}{0} \ \A{\nu'}{b'}{0}: \  + \
\dots \;\; ; \nonumber \\
\label{OPEfield2}
\eeq

\noindent where only  normal products of local gluon  field
operators occur and $A$ (${\widetilde A}$) stands for the gluon field in 
configuration (momentum) space, $a,b$ being colour indices and $\mu,\nu$ Lorentz
ones. The notation $T()$ simply refers to the standard $T^\ast$ product  in
momentum space. The normal product of Eqs. (\ref{OPEfield1},\ref{OPEfield2})
should be defined in reference to the perturbative vacuum~\cite{OPE}. 

\begin{figure}
%\vspace*{-1.cm}
\hspace*{-1.3cm}
\begin{center}
%\begin{tabular}{ll}
%\epsfxsize5.5cm\epsffile{OPEpropagA.eps} & 
\epsfxsize7.5cm\epsffile{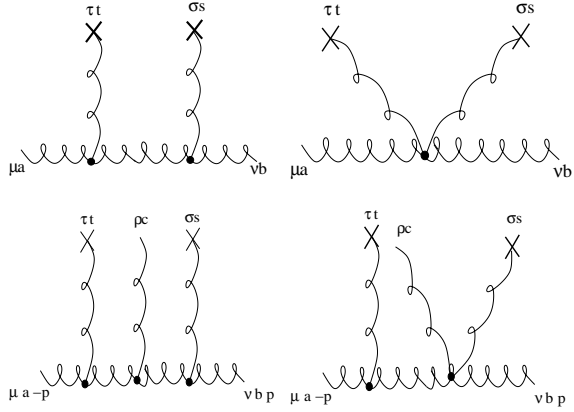} %\\
%\end{tabular}
\caption{\small {\it Four and five gluons tree-level diagrams contributing (with all their 
possible permutations) to the Wilson coefficients of the gluon 
propagator and the three-gluon vertex. Crosses mark the gluon legs due to the external
soft gluons.}}
\label{Fig1}
\end{center}
\end{figure}

Only terms in Eqs. (\ref{OPEfield1},\ref{OPEfield2})  containing an even number
of local gluon fields give a non-null {\it. v.e.v.} because of Lorentz invariance and of the gauge
condition~\cite{OPE}. The coefficients $c_0$ and $d_0$ are the purely 
perturbative Green functions. Assuming the Wilson factorization of  soft and hard
gluon  contributions, the relevant Wilson coefficients $c_2,d_2$ can be obtained,
in perturbation, by computing the diagrams  in Fig. \ref{Fig1} which represent
the matrix elements of operators  in the l.h.s. of Eqs.
(\ref{OPEfield1},\ref{OPEfield2}) between soft gluons, indicated by crosses.
Using also the tree level expression for the matrix element, between  the same
two soft gluons, of the local operators in the r.h.s. of  Eqs.
(\ref{OPEfield1},\ref{OPEfield2}) we obtain $c_2$ and $d_2$ by matching both
sides. 

%%%%%%%%%%%%%%%%%%%%%%%%%%%%%%%%%%%%%%%%%%%%%%%%%%%%%%%%%%%%%%%%%%%%%%
%%%%%%%Figure 1
%%%%%%%%%%%%%%%%%%%%%%%%%%%%%%%%%%%%%%%%%%%%%%%%%%%%%%%%%%%%%%%%%%%%%%
%%\begin{figure}[hbt]
%%\begin{center}
%%\leavevmode
%%\epsfysize=11.0truecm
%%\epsffile{ZAtot_orbites.eps}
%%\epsfysize=11.0truecm
%%\epsffile{ZAcompare.eps}
%%\mbox{\epsfig{file=Fig1a.eps,height=8cm}}
%%\mbox{\epsfig{file=Fig1b.eps,height=8cm}}
%%\vspace*{-4.5cm}
%%%%%%%%%%%%%%%%%%%%%%%%%%%%%%%%
%%%%%%%%%%%%%%%%%%%%%%%%%%%%%%%%%%%%%%%%%%%%%%%%%%%%%%%%%%%%%%%%%%%%%%%

Thus, in the appropriate Euclidean metrics for matching to lattice
non-perturbative results,  we can write:

\beq
k^2 G^{(2)}(k^2) & = & Z^{MOM}(k^2) \ = \ Z^{MOM}_{\rm n loops}(k^2) \ + 
\frac{3g^2\langle A^2 \rangle}{4(N_c^2-1)} \ \frac{1}{k^2}  \; ,
\nonumber \\
k^6 G^{(3)}(k^2,k^2,k^2) & = & k^6 G^{(3)}_{\rm pert}(k^2,k^2,k^2) \ + \ 
\frac{9g^3\langle A^2 \rangle}{4(N_c^2-1)} \ \frac{1}{k^2}  \; ;
\label{OPEscalars}
\eeq

\noindent where the scalar form factors $G^{(2)},G^{(3)}$ are defined  as follows
from the Green functions 

\beq
G^{(2)}(p^2) &=& 
\frac{\delta_{a b}}{N_c^2-1}\; \frac{1}{3}
\left(\delta_{\mu \nu}-\frac{p_\mu p_\nu}{p^2}\right) \ 
\langle \Am{\mu}{a}{-p} \ \Am{\nu}{b}{p} \rangle \; , 
\nonumber \\
G^{(3)}(k^2,k^2,k^2) & = & \frac{1}{18k^2} \ \frac{f^{a b c}}{N_c (N_c^2-1)}
\langle \Am{\mu}{a}{p_1} \Am{\nu}{b}{p_2} \Am{\rho}{c}{p_3}\rangle \nonumber \\
&\times & \left[(T^{\rm tree})^{\mu_1 \mu_2 \mu_3}+\frac{(p_1-p_2)^\rho (p_2-p_3)^\mu 
(p_3-p_1)^\nu}{2k^2} \right] \; .
\label{EucGreens}
\eeq

\noindent For the kinematic configuration $p_1^2=p_2^2=p_3^2=k^2$ 
 the three-gluon tree-level tensor is defined as

\beq
(T^{\rm tree})_{\mu_1 \mu_2 \mu_3}=\left[ \delta_{\mu'_1 \mu'_2} (p_1-p_2)_{\mu'_3} 
+ \rm{cycl. perm.} \right] \ \prod_{i=1,3} \left( \delta_{\mu'_i \mu_i}-
\frac{p_{i \mu'_i} p_{i \mu_i}}{p_i^2} \right) . \nonumber \\
\label{Ttree}
\eeq

In Eqs. (\ref{OPEscalars})-(\ref{Ttree}) we have dealt with bare quantities,
depending only on the  cut-off $a^{-1}$ and on the momentum $k$.
We have omitted to explicitate the dependence on the cut-off
in order to simplify the notations. 
Using Eqs. (\ref{OPEscalars}) these Green functions can be conveniently 
renormalized by MOM prescriptions: the renormalized two-point Green  function 
is taken equal to $1/k^2$ for $k^2=\mu^2$,

\beq
k^2 G^{(2)}_R(k^2,\mu^2)  \equiv  \frac {k^2 G^{(2)}(k^2)}{\mu^2 G^{(2)}(\mu^2)} =
c_0\left(\frac{k^2}{\mu^2},
\alpha(\mu)\right) \ + 
c_2\left(\frac{k^2}{\mu^2},\alpha(\mu) \right) \ 
\frac{\langle A^2 \rangle_{R,\mu}}{4(N_c^2-1)} \ \frac{1}{k^2} \; ,\label{gren}
 \eeq

\noindent The $c_0$ Wilson coefficient can be written as 

\beq
c_0 \left(\frac{k^2}{\mu^2},\alpha(\mu) \right) \ = \ 
\frac{Z_{\rm nloops}^{\rm MOM}(k^2)}{Z^{\rm MOM} (\mu^2)} \ = \ 
c_0 \left(1,\alpha(\mu)\right) 
\frac{Z_{\rm nloops}^{\rm MOM}(k^2)}{Z_{\rm nloops}^{\rm MOM}(\mu^2)} \; \; ,
\eeq

\noindent and verifies consequently the perturbative evolution equations of 
$Z^{\rm MOM}$, 

\beq
\frac{d \ln c_0\left(\frac{k^2}{\mu^2},
\alpha(\mu)\right)}{d \ln{k^2}}  =  
- \ \left(
\gamma_0 \frac{\alpha(k)}{4 \pi} \ + \ \gamma_1 \left(\frac{\alpha(k)}{4 \pi}\right)^2
\ + \ \gamma_2 \left(\frac{\alpha(k)}{4 \pi}\right)^3 \ + \ \dots \ \right) 
%\nonumber \\
%& = & 
%- \ \left(
%\gamma_0 \frac{\widetilde{\alpha}(k)}{4 \pi} \ + \ \widetilde{\gamma_1} 
%\left(\frac{\widetilde{\alpha}(k)}{4 \pi}\right)^2
%\ + \ \widetilde{\gamma_2} \left(\frac{\widetilde{\alpha}(k)}{4 \pi}\right)^3 \ 
%+ \ \dots \; \right) , \ \ 
\label{ZMOMR}
\eeq

\noindent where $\gamma_1$ and $\gamma_2$ depend on the perturbative
scheme in which the strong coupling constant $\alpha(k)$ is defined. The 
boundary condition to solve Eq. (\ref{ZMOMR}) comes 
from the nonperturbative normalization of 
$k^2 G^{(2)}_R(k^2,\mu^2)$ to 1 at $k^2=\mu^2$, and it results that 
$c_0(1,\alpha(\mu)) = 1 + O(1/\mu^2)$. 
 
Let us remind that in the MOM prescription, the three-point Green function 
is renormalized by $G_R^{(3)}(k^2,\mu^2) \equiv 
G^{(3)}(k^2,k^2,k^2) (Z^{\rm MOM}(\mu))^{-3/2}$, and the MOM coupling constant 
follows from

\beq
g_R(k^2)  & = & \frac{G^{(3)}(k^2,k^2,k^2)}{\left(G^{(2)}(k^2)\right)^3} \ 
\left(Z^{\rm MOM}(k^2)\right)^{3/2} \nonumber \\
& = & k^6 G^{(3)}_R(k^2,\mu^2) \left( k^2 G^{(2)}_R(k^2,\mu^2) \right)^{-3/2} \; \; .
\label{gr}
\eeq

Analogously to Eq. (\ref{gren}) we define 
the renormalized three point Green function

\beq 
 k^6 G^{(3)}_R(k^2,\mu^2) =
d_0\left(\frac{k^2}{\mu^2},\alpha(\mu)\right) \ + \ 
d_2\left(\frac{k^2}{\mu^2},\alpha(\mu)\right) \
\frac{\langle A^2 \rangle_{R,\mu}}{4(N_c^2-1)} \ \frac{1}{k^2}  \; .
\label{OPEescalarsR}
\eeq
where the $d_0$ Wilson coefficient verifies  the perturbative
evolution equations of $k^6 G^{(3)}_R(k^2,\mu^2)$ and the boundary condition 
$d_0(1,\alpha(\mu)) = g_R(\mu^2) + O(1/\mu^2)$ is immediate from 
Eqs. (\ref{gr},\ref{OPEescalarsR}).
 
In the MOM scheme, the gluon condensate, $\langle A^2 \rangle_{R,\mu}$, 
is renormalized at $\mu^2$ by a standard condition, through division by a 
renormalization constant $Z_{A^2}$.

The $c_2$ and $d_2$ Wilson coefficients at tree level are~\cite{OPE}  

\beq
c_2\left(1,\alpha(\mu) \right)=3 \ g^2 \; \; , \nonumber \\
d_2\left(1,\alpha(\mu) \right)=9 \ g^3 \; \; . \label{treecoef}
\eeq

Since the three point Green function naturally defines the MOM
scheme coupling constant (see below (\ref{gr})), we will perform 
all the coming calculations in the symmetric MOM scheme where 
\beq
\gamma_0 = 13/2, \quad \gamma_1=-16.9, \quad 
\gamma_2 \simeq 1332.3 \; \; .
\eeq
 
\noindent  In Eq. (\ref{ZMOMR})  $\alpha(k)$ is of course taken  to be the
purely perturbative running coupling  constant, $g_{\rm R,pert}^2(k^2)/(4 \pi)$,
obtained by integrating the beta function  in  the MOM scheme,

\beq
\frac{d}{d\ln{k}} \ \alpha(k) \ = \ 
\beta\left(\alpha(k)\right) \ = \
- \left( \frac{\beta_0}{2 \pi} 
\alpha^2(k) \ + \ \frac{\beta_1}{4 \pi^2} \ \alpha^3(k)
\ + \ \frac{\beta_2}{(4 \pi)^3} \ \alpha^5(k) \ 
+ \ \dots \ \right) \ ,
\label{beta}
\eeq

\noindent where~\cite{Chety2}

\beq
\beta_0 = 11 , \quad \beta_1 = 51 , \quad \beta_2 \simeq 3088. \; \; .
\label{betaMOM}
\eeq

%=============================================
\subsection{Wilson coefficient at leading logs}
\label{wilson}
\hspace*{\parindent} The purpose is now to compute to leading logarithms the subleading Wilson
coefficients in  Eqs. (\ref{gren},\ref{OPEescalarsR}). To this goal, 
following~\cite{Rafael} it will  be useful to consider the following matrix
element,

\beq \langle g^a_\tau | \Am{\rho}{r}{k} \Am{\sigma}{s}{-k} | g^b_\nu
\rangle_{R,\mu} \ = \ \delta^{r s} \left( \delta_{\rho \sigma}-\frac{k_\rho k_\sigma}{k^2} \right) \ 
\left[\frac{c_2\left(\frac{k^2}{\mu^2},\alpha(\mu)\right)}{k^4} \  \frac{\langle
g^a_\tau | A^2_R | g^b_\nu \rangle_{\mu}}{4 (N_C^2-1)} \ + \ \dots \ \right]
\ , \label{R} \eeq

\noindent where the external gluons carry soft momenta. Dots inside the
brackets refer  to terms with powers of $1/k$ different from $4$ (i.e.
corresponding to higher dimension operators or to identity
operator~\footnote{It should be  remembered that other terms,  like
$\partial_\mu A_\mu$, with the same dimension of $A^2$,  do not survive}).  
From eq. (\ref{R}) we get 

\beq 4 (N_C^2-1) k^4 \ \frac{\langle g^a_\tau | \Am{\rho}{r}{k}
 \Am{\sigma}{s}{-k} | g^b_\nu
\rangle}{ \langle g^a_\tau | A^2 | g^b_\nu \rangle} \  \delta^{r s} \left(
\delta_{\rho \sigma}-\frac{k_\rho k_\sigma}{k^2} \right) \  & = & \
Z_3(\mu^2)Z_{A^2}^{-1}(\mu^2) \ c_2\left(\frac{k^2}{\mu^2},\alpha(\mu)\right) \
+  \dots  \nonumber \\ & \equiv & \ \widehat{Z}^{-1}(\mu^2) \
c_2\left(\frac{k^2}{\mu^2},\alpha(\mu)\right) \  + \dots \ ,  \label{b} \eeq

\noindent where $\widetilde{A}_R=Z_3^{-1/2} \widetilde{A}$ and
$A^2_R=Z_{A^2}^{-1} A^2$,  while $\widehat{Z}\equiv Z_3^{-1} Z_{A^2}$ is a
useful notation denoting  the divergent factor of the matrix element coming
from proper vertex corrections. If one takes the  logarithmic
derivatives with respect to  $\mu$ in both  Eq. (\ref{b})'s hand-sides,
the following differential equation is obtained:
\beq
\left\{ \ -2 \gamma\left(\alpha(\mu)\right)_{A^2}+2 \gamma\left(\alpha(\mu)\right)
+\frac{\partial}{\partial\ln{\mu}} +\beta\left(\alpha(\mu)\right)  
\frac{\partial}{\partial\alpha} \ \right\} \ c_2\left(\frac{k^2}{\mu^2},\alpha(\mu)\right) 
\ = \ 0 \ ;
\label{RG1}
\eeq

An analogous differential equation describing the behaviour of the three-point 
Wilson coefficient on the renormalization momentum, $\mu$, can be 
obtained similarly,

\beq
\left\{ \ -2 \gamma\left(\alpha(\mu)\right)_{A^2}
+3 \gamma\left(\alpha(\mu)\right) + 
\frac{\partial}{\partial\ln{\mu}}+\beta\left(\alpha(\mu)\right)  
\frac{\partial}{\partial\alpha} \ \right\} \ d_2\left(\frac{k^2}{\mu^2},\alpha(\mu)\right) 
\ = \ 0 \ .
\label{RG2}
\eeq
\noindent Here we have defined :
\beq
\gamma_{A^2}=\frac{d}{d\ln{\mu^2}} \ln{Z_{A^2}}
\eeq
\noindent and $\gamma\left(\alpha(\mu)\right)$ is the gluon propagator 
anomalous dimension. 
Reexpressing these evolution equations in terms of 
\beq
\widehat{\gamma}\left(\alpha(\mu)\right) \ = \ 
\frac{d}{d\ln{\mu^2}} \ln{\widehat{Z}(\mu^2)} \ .
\label{Ad}
\eeq
We obtain :
\beq
\left\{ \ -2 \widehat{\gamma}\left(\alpha(\mu)\right)
+\frac{\partial}{\partial\ln{\mu}}+\beta\left(\alpha(\mu)\right)  
\frac{\partial}{\partial\alpha} \ \right\} \ c_2\left(\frac{k^2}{\mu^2},\alpha(\mu)\right) 
\ = \ 0 \ ;
\label{Rafprop1}
\eeq
and 
\beq
\left\{ \ -2 \widehat{\gamma}\left(\alpha(\mu)\right)
+\gamma\left(\alpha(\mu)\right) + 
\frac{\partial}{\partial\ln{\mu}}+\beta\left(\alpha(\mu)\right)  
\frac{\partial}{\partial\alpha} \ \right\} \ d_2\left(\frac{k^2}{\mu^2},\alpha(\mu)\right) 
\ = \ 0 \ .
\label{Rafprop2}
\eeq

The leading log solution for both Eqs. (\ref{Rafprop1},\ref{Rafprop2})
can be written as,

\beq
c_2\left(\frac{k^2}{\mu^2},\alpha(\mu)\right) & = & c_2\left(1,\alpha(k)\right) \ 
\left( \frac{\alpha(k)}{\alpha(\mu)} 
\right)^{- \frac{\widehat{\gamma}_0} {\beta_0}} \ , \nonumber \\
d_2\left(\frac{k^2}{\mu^2},\alpha(\mu)\right) & = & d_2\left(1,\alpha(k)\right) \
\left( \frac{\alpha(k)}{\alpha(\mu)} 
\right)^{\frac{\gamma_0 - 2 \widehat{\gamma}_0} {2 \beta_0}} \ .
\label{SolRaf}
\eeq
\noindent where $\widehat{\gamma}_0$ is defined in analogy with $\gamma_0$ :
\beq
\widehat{\gamma}\left(\alpha(\mu)\right) = - \widehat{\gamma}_0 \frac{\alpha(k)}{4 \pi} 
+ ... 
\eeq

%We still need to estimate  the prefactors $c_2\left(1,\alpha(k) \right)$ and 
%$d_2\left(1,\alpha(k) \right)$. We are free to choose, in the spirit of the MOM
%scheme, 
%\beq
%c_2\left(1,\alpha(k) \right)=3 \ g^2_R(k^2) \; \; ,\label{treec} 
%%\eeq 
%\noindent as is natural in view of the tree level result (\ref{treecoef}).
%This choice {\it defines} the normalization of the $A^2$
%condensate.

The prefactors  $c_2\left(1,\alpha(k)\right) $ and  $d_2\left(1,\alpha(k)\right) $  have to be matched at tree level to Eq.~(\ref{treecoef}). The only solutions are of the form :

\beq
c_2\left(1,\alpha(k) \right)=3 \ g_R^2(k^2)\left ( 1 + {\cal O}\left (\frac{1}{\log(k/\Lambda_{\rm QCD})}\right) \right)  \; \; , \nonumber \\
d_2\left(1,\alpha(k) \right)=9 \ g_R^3(k^2) \left ( 1 + {\cal O}\left (\frac{1}{\log(k/\Lambda_{\rm QCD})}\right) \right)\; \; . 
\eeq

The ${\cal O}\left(\frac{1}{\log(k/\Lambda_{\rm QCD})}\right)$ terms are clearly of the same order as the next-to-leading contributions to the anomalous  dimension which are systematically omitted in this paper.

Of course, these solutions of Eqs. (\ref{Rafprop1},\ref{Rafprop2}) define the dependence of the Wilson coefficients  not
only on the renormalization momentum,
$\mu$,  but simultaneously  on the momentum scale $k^2$. This is a 
straightforward consequence  of standard dimensional arguments: the only
dimensionless  quantities are the ratio $k^2/\mu^2$  and $\alpha$.  Then, as soon
as one knows perturbatively  $\widehat{\gamma}\left(\alpha(\mu)\right)$, 
$\gamma\left(\alpha(\mu)\right)$ and $\beta\left(\alpha(\mu)\right)$, the leading
logarithmic behaviour on $k$ is available.

 As already mentioned, the gluon propagator  anomalous dimension and the beta
function are  known up to three loops in  the MOM scheme 
and up to three and four loops,  respectively, in the $\widetilde{\rm MOM}$.  The anomalous dimension
of the $A^2$  operator is obviously less stimulating for  perturbative QCD
community. We have done this calculation to one loop  (see appendix), obtaining

\beq
\widehat{\gamma}\left(\alpha(\mu)\right) \ = 
\ - \widehat{\gamma}_0  \ \frac{\alpha(\mu)}{4 \pi} +...\ =
\ - \frac{3 N_C}{4} \ \frac{\alpha(\mu)}{4 \pi} +... \ \ .
\label{gA20}
\eeq

and 
\beq \gamma_{A^2}\left(\alpha(\mu)\right)=\frac{d}{d\ln{\mu^2}} \ln{Z_{A^2}}
=-{35 N_C \over 12}\frac{\alpha(\mu)}{4 \pi} +... \eeq

%================================================
\subsection{Gluon propagator with leading logs for the condensate coefficient} 

\hspace*{\parindent} Let us now specify our approach to lattice results.
Using the definitions in Eqs. (\ref{OPEscalars}) and (\ref{gren})
we will match our lattice results to  

\beq
\frac{Z^{\rm MOM}_{\rm Latt}(k^2,a)}{Z^{\rm MOM}_{\rm Latt}(\mu^2,a)} \ = \ 
k^2 G^{(2)}_R(k^2,\mu^2) \ + \ O(a^2) \ ,
\label{ZLattInd}
\eeq

\noindent where the adequate control of lattice artifacts\footnote{See ref. 
\cite{propag}, where we discuss at length the artifacts of the lattice gluon 
propagator evaluation.} reduces the UV discretization errors to an acceptable
level.  From Eq. (\ref{gren}),

\beq
k^2 G^{(2)}_R(k^2,\mu^2)  & = & 
c_0\left(\frac{k^2}{\mu^2},\alpha(\mu)\right) \nonumber \\ 
& \times & \left( \ 1 \ +  
\frac{c_2\left(\frac{k^2}{\mu^2},\alpha(\mu) \right)}
{c_0\left(\frac{k^2}{\mu^2},\alpha(\mu)\right)} \ 
\frac{\langle A^2 \rangle_{R,\mu}}{4(N_c^2-1)} \ \frac{1}{k^2} \right) \; 
\label{Z0};
\eeq

\noindent where we explicitly factorize the  Wilson  coefficient of the
identity operator which, as was previously indicated, is known to three loops. 
Nevertheless, for consistency, all the terms {\it inside the parenthesis} in the
r.h.s of Eq.(\ref{Z0})  will be developed only to leading order, including  
$c_0\left(\frac{k^2}{\mu^2},\alpha(\mu)\right)$,

\beq
c_{0,{\rm LO}}\left(\frac{k^2}{\mu^2},\alpha(\mu)\right) \, = \,\left (\frac{\alpha(k)}{\alpha(\mu)}\right)^{\frac{\gamma_0}{\beta_0}} 
%& = & \left( \frac{\ln{\frac{k}{\Lambda}}}{\ln{\frac{\mu}{\Lambda}}} 
%\right)^{-\frac{\gamma_0}{\beta_0}} + O(1/\mu^2) \; .
\label{ZR1loop}
\eeq

\noindent Terms of the order of $O(1/(k^2 \mu^2) )$ have been neglected, as well as, of course, 
those of $O(1/k^4)$ coming from higher dimension operators.
 One free parameter, i.e. a boundary condition, has to be fitted from
lattice data. It can be   either $\alpha(\mu)$ or the  $\Lambda$ parameter, i.e.
the position of the perturbative  Landau pole. We choose  the latter.  We write $c_{0,{\rm 1
loop}}$ in terms of the MOM coupling constant\footnote{$\widetilde{\rm MOM}$, for
instance, or whatever renormalization  scheme could be used alternatively.
Our preference for the MOM scheme has been explained above.} and the
$\Lambda$ parameter in Eq. (\ref{ZR1loop}) 
in the MOM scheme\footnote{See, for instance, ref. \cite{Alles}},

\beq
\Lambda \equiv \Lambda^{\rm MOM} \simeq 3.334 \ \Lambda^{\overline{\rm MS}} \; .
\eeq

\noindent We finally obtain 

\beq
Z^{\rm MOM}_{\rm Latt}(k^2,a) & = &
Z^{\rm MOM}_{\rm Latt}(\mu^2,a) \ 
c_0\left(\frac{k^2}{\mu^2},\alpha(\mu)\right) \nonumber \\
& \times & \left( 1 \ + \ R^{(2)} \ 
\left( \ln{\frac{k}{\Lambda}}\right)^{\frac{\gamma_0 + 
\widehat{\gamma}_0}{\beta_0}-1} \ \frac{1}{k^2} \ \right) \; ;
\label{ZLatt}  
\eeq

\noindent where

\beq
R^{(2)} \ = \ \frac{6 \pi^2}{\beta_0 \ (N_c^2-1)} \ 
\left( \ln{\frac{\mu}{\Lambda}} 
\right)^{-\frac{\gamma_0 + 
\widehat{\gamma}_0}{\beta_0}} \langle A^2 \rangle_{{\rm R},\mu} \; .
\label{R2}
\eeq

%======================================
\subsection{Running coupling constant} 

\hspace*{\parindent} %Calling $g_{\rm Latt}(k,a)$ the running coupling constant in the MOM scheme,  Eq. (\ref{gr}),
 %issuing from lattice results we have

%\beq
%g_{\rm Latt}(k^2,a)=k^6 G_R^{(3)}(k^2,\mu^2) \left(k^2 G_R^{(2)}(k^2,\mu^2)\right)^{-3/2}
%\ + \ O\left((ak)^2\right) \; ;
%\label{gLdeka}
%\eeq

%\noindent where we checked (cf \cite{poweral} where wee also explain how our data set was built) that the UV lattice artifacts are under %control by 
%testing the scaling  on lattices with different spacings, $a$.
 By taking the OPE expansions in Eqs. (\ref{OPEescalarsR}) and  
(\ref{gren})  (\ref{gr}) can be written as

\beq
g_{\rm R}(k^2) & = & g_{\rm R, pert}(k^2) \ \left\{ 1 \ + \
\frac{\langle A^2 \rangle_{R,\mu}}{4(N_c^2-1)} \ \frac{1}{k^2} \ 
\left( \frac{ d_2\left(\frac{k^2}{\mu^2},\alpha(\mu) 
\right)}{\left[ d_0\left( 
\frac{k^2}{\mu^2},\alpha(\mu) \right) \right]} \right. \right. 
\nonumber \\
& - & \left. \left. \frac{3}{2} \ \frac{ c_2\left(\frac{k^2}{\mu^2},\alpha(\mu) 
\right)}{c_0\left( 
\frac{k^2}{\mu^2},\alpha(\mu) \right)}   \right) \right\} \; ;
\label{gfin}
\eeq

\noindent with the identification

\beq
g_{\rm R, pert}(k^2) \ \equiv \ d_0\left(\frac{k^2}{\mu^2},\alpha(\mu) \right) \   
\left[ c_0\left(\frac{k^2}{\mu^2},\alpha(\mu) \right) \right]^{-3/2} \; ,
\label{gpert}
\eeq

Applying the results given by Eqs. (\ref{SolRaf},\ref{ZLatt}--\ref{gpert}), 
 $\alpha_{\rm MOM}=g_{\rm R}^2/(4 \pi)$ verifies

\beq
\alpha_{\rm MOM}(k) \ = \ \alpha_{\rm pert}(k) \ \left( 1 \ + \ 
R^{(3)} \ 
\left( \ln{\frac{k}{\Lambda}}\right)^{\frac{\gamma_0 + 
\widehat{\gamma}_0}{\beta_0}-1} \ \frac{1}{k^2} \ 
\right) \; ;
\label{alphaLatt}
\eeq

\noindent where 

\beq
R^{(3)} \ = \ \frac{18 \pi^2}{\beta_0 \ (N_c^2-1)} \ 
\left( \ln{\frac{\mu}{\Lambda}} 
\right)^{-\frac{\gamma_0 + 
\widehat{\gamma}_0}{\beta_0}} \langle A^2 \rangle_{{\rm R},\mu} \; .
\label{R3}
\eeq

\noindent Again, we do not retain $O(1/(k^2 \mu^2),1/k^4)$-terms.

%**************************************
\section{Fitting the data to our ans\"atze} 
\label{sec:lattice}

\hspace*{\parindent} We shall follow in this section the OPE testing approach proposed in ref. 
\cite{OPE}:  trying a consistent description of lattice data for two- and
three-gluon Green functions from ref. \cite{frenchalpha,propag,poweral}.  We are
however in a much better position than in \cite{OPE}.  In the latter work, only 
two-loop information was available for the beta function  and $d_0$ while the
subdominant Wilson coefficients $c_2$ and $d_2$ were computed only at
tree-level.  This had the practical inconvenience of preventing a simultaneous
fit of both   $\Lambda^{\overline{\rm MS}}$ and $\langle A^2 \rangle$: the
$\Lambda^{\overline{\rm MS}}$ parameter had to be taken from outside our matching
procedure. Now the
new input for three-loop MOM beta function and $c_0$ coefficient~\cite{Chety2} enable us to perform a self-consistent test
by combining the matching  of the  gluon propagator and of $\alpha_s^{\rm MOM}$ to
formulas in Eqs.  (\ref{ZLatt}-\ref{R2},\ref{alphaLatt}-\ref{R3}),  where the
three quantities, $\Lams$ and gluon condensates from both Green functions, are
taken  to be fitted on the same footing. Of course, the test consists in checking
the equality of the two gluon condensates obtained from those two different
 Green functions. 

In the case of the gluon propagator, the factor  
 $c_0(1,\alpha(\mu)) Z^{\rm MOM}_{\rm Latt}(\mu^2,a)$,
  which carries all the logarithmic dependence on the  lattice
spacing, appears as an additional parameter to be fitted.  As we explained in
ref. \cite{OPE}, a large fitting window is an  important ``{\it ace}'' to
restrict the potentially dangerous confusion between Wilson coefficients for
different powers. To combine data over such a large energy window we need to
match the lattice  results obtained with different lattice spacings and the
last factor carrying lattice spacing  dependence should be independently
fitted for each one. On the  contrary, the running  coupling constant should
be regularization independent and the matching of data sets corresponding to different lattice spacing
 can be imposed without tuning any additional parameter (this  is by itself a
positive test  of the goodness of the  procedure used to build our data set).  As a
matter of fact, this is why the matching of the latter to perturbative
formulas  is much more constraining than that of the former in order to
estimate $\Lams$, as  discussed in refs. \cite{frenchalpha,propag,poweral}.
The details of the lattice simulations, of the procedures used to
obtain an artifact-safe data set or of the definition of
regularization-independent objects permitting lattice regularized data to be
matched to continuum quantities in any scheme, can be found in those
references. We will now present the results of the fitting strategy just
described.

\paragraph{Two loop fit}

\begin{figure}[hbt]
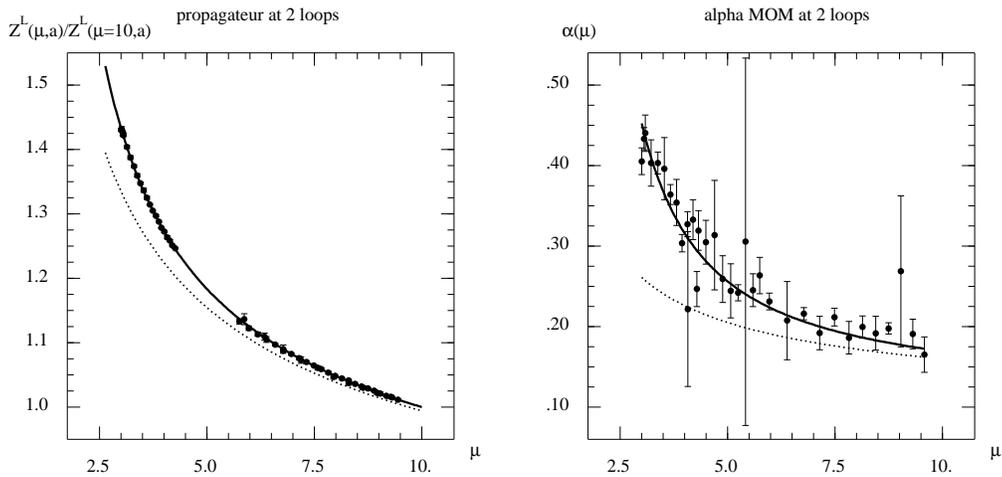

%\vspace*{-1.cm}
\hspace*{-1.3cm}
\begin{center}
\begin{tabular}{ll}
\epsfxsize6.5cm\epsffile{prop_2boucles.eps} & 
\epsfxsize6.5cm\epsffile{alpha_2boucles.eps} \\
\end{tabular}
\caption{\small {\it Comparison of the 2-loops fit to the ratio of the renormalization constants at $k$ and at $10\, GeV$ 
 and to $\alpha_s(k)$ with the lattice data for $2.5 <k < 10\, GeV$.The dotted line shows the perturbative part.}}
\label{Fig2}
\end{center}
\end{figure}

We first perform the combined fit for two- ant three-gluon Green function  at
two-loop level for the leading Wilson coefficients. In fig. \ref{Fig2}, we plot
lattice  data and the curves given by Eqs.
(\ref{ZLatt}-\ref{R2},\ref{alphaLatt}-\ref{R3}) with  the following best-fit
parameter:

\beq
&&{\rm propagator :}\quad \sqrt{\langle A^2 \rangle_{R,\mu}} = 1.64(17) 
{\rm GeV};\qquad {\alpha^{\rm MOM} :}
\quad\sqrt{ \langle A^2 \rangle_{R,\mu}} = 3.1(3) {\rm GeV}.
\nonumber \\
&&\frac{\left\{\sqrt{\langle A^2 \rangle_{R,\mu}}\right\}_{\rm alpha}}
{\left\{\sqrt{\langle A^2 \rangle_{R,\mu}}\right\}_{\rm prop}} \ = \ 1.86(4); 
\qquad \Lams = 172(15) {\rm MeV} 
\label{twoloops}
\eeq 

\noindent with a $\chi^2/d.o.f \simeq 1.1$ for the combined fit. 

%%%%%%%%%%%%%%%%%%%%%%%%%%%%%%%%%%%%%%%%%%%%%%%%%%%%%%%%%%%%%%%%%%%%%
%%%%%%Figure 2
%%%%%%%%%%%%%%%%%%%%%%%%%%%%%%%%%%%%%%%%%%%%%%%%%%%%%%%%%%%%%%%%%%%%%
%%%%%%%%%%%%%%%%%%%%%%%%%%%%%%%%%%%%%%%%%%%%%%%%%%%%%%%%%%%%%%%%%%%%%%%

\paragraph{Three loops fit}

\begin{figure}[hbt]
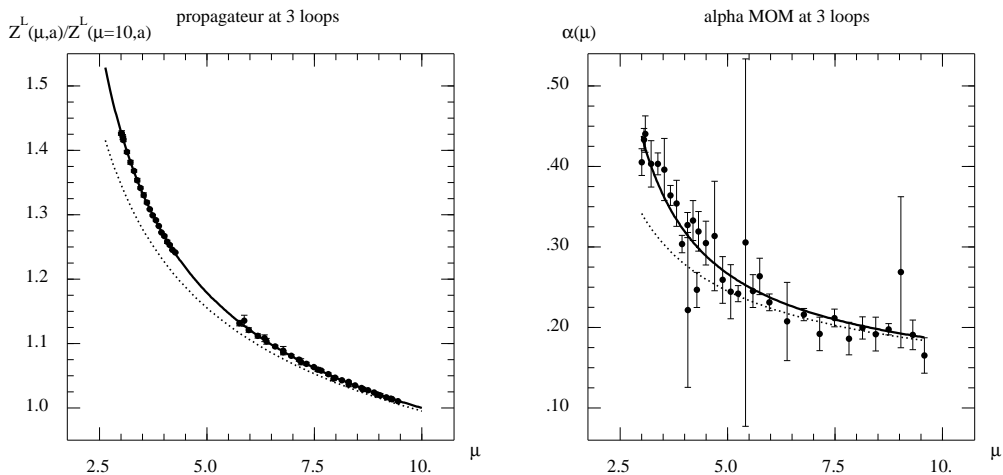

%\vspace*{-1.cm}
\hspace*{-1.3cm}
\begin{center}
\begin{tabular}{ll}
\epsfxsize6.5cm\epsffile{prop_3boucles.eps} & 
\epsfxsize6.5cm\epsffile{alpha_3boucles.eps} \\
\end{tabular}
\caption{\small {\it Same as Figure 2 at 3-loops level.}}
\label{Fig3}
\end{center}
\end{figure}

The present perturbative knowledge allows a three-loop level fit for
leading Wilson coefficients. Analogously  to the previous paragraph, 
we plot in Fig.
\ref{Fig3} the lattice  data and curves given by  Eqs.
(\ref{ZLatt}-\ref{R2},\ref{alphaLatt}-\ref{R3}), $Z_{\rm R, pert}^{\rm MOM}$ and
$\alpha_{\rm pert}$ taken at three loops, with the following best-fit
parameters:

\beq
&&{\rm propagator :}\quad \sqrt{\langle A^2 \rangle_{R,\mu}} = 1.55(17) {\rm GeV};\qquad
{\alpha^{\rm MOM} :}
\quad \sqrt{\langle A^2 \rangle_{R,\mu}} = 1.9(3) {\rm GeV}.
\nonumber \\
&&\frac{\left\{\sqrt{\langle A^2 \rangle_{R,\mu}}\right\}_{\rm alpha}}
{\left\{\sqrt{\langle A^2 \rangle_{R,\mu}}\right\}_{\rm prop}} \ = \ 1.21(18); 
\qquad \Lams = 233(28) {\rm MeV} 
\label{threeloops}
\eeq 

\noindent with $\chi^2/d.o.f \simeq 1.2$. Combining the results obtained from $\alpha^{\rm MOM}$
 and from the propagator in the standard way gives our final result $\sqrt{\langle A^2 \rangle_{R,\mu}} = 1.64(15) {\rm GeV}$.
The renormalization scale $\mu$ is taken to be $10$ GeV in both combined  fits at
two- and three-loop level. However we have checked that, varying $\mu$  over the
fitting window where we can legitimately neglect terms of  $O(1/(\mu^2 k^2))$ in
Eq. (\ref{R3}), the ratios of condensates in Eqs. 
(\ref{twoloops},\ref{threeloops}) remain essentially unmodified. In fact, that 
$R^{(2)}$ in Eq. (\ref{ZLatt}) does not depend on $\mu$  has been  explicitly
tested over the fitting window (the same is obvious for  $R^{(3)}$ in Eq.
(\ref{alphaLatt}) where nothing depends on $\mu$). %It is also important to notice
%that   the jackknife errors for the ratios turn out to be considerably smaller that the ones
%for the  fitted condensates. This clearly indicates the correlated nature of
%these last  condensates' errors, estimated by jacknife methods, from the combined  fitting procedure described above.

%%%%%%%%%%%%%%%%%%%%%%%%%%%%%%%%%%%%%%%%%%%%%%%%%%%%%%%%%%%%%%%%%%%%%
%%%%%%Figure 3
%%%%%%%%%%%%%%%%%%%%%%%%%%%%%%%%%%%%%%%%%%%%%%%%%%%%%%%%%%%%%%%%%%%%%
%%%%%%%%%%%%%%%%%%%%%%%%%%%%%%%%%%%%%%%%%%%%%%%%%%%%%%%%%%%%%%%%%%%%%%%

\section{Discussion and conclusions}
\protect\label{sec:conclusions}
\hspace*{\parindent}The gluon condensate $\langle A^2 \rangle$ has  been 
computed from the
deviations of both the lattice
non-perturbative evaluation of the MOM $\alpha_S$  and the gluon propagator from their known perturbative behaviour. We
have described these non-perturbative deviations using  OPE, and  fitting the
condensate  to match both sides. The use of the self-consistent fitting
strategy described  in the preceding section leads  simultaneously to a
prediction $\Lams$ and  to two independent  estimates of $\langle A^2 \rangle$.

%resulting from two- and three-point Green %function. 

The fit using the two-loops perturbative expressions for both the MOM $\alpha_S$
and the gluon propagator  clearly fails: {\it a clear disagreement between the
two independent estimates of  $\sqrt{\langle A^2 \rangle}$ is found}. The ratio
of both estimates is $1.86(4)$ from Eqs. (\ref{twoloops}). This confirms the
preliminary analysis in ref. \cite{OPE}, where  only tree-level Wilson
coefficients were computed.% 
%As a matter of fact, this conclusion is clearly confirmed by the  analysis with
%three-loop perturbative MOM $\alpha_S$ and gluon propagator. 
In this preliminary work, a self-consistent three-loop analysis was not possible because the  MOM beta function was not known up to three-loops. 
Nevertheless we  tried to fit the third coefficient of the beta function,
$\beta_2$ to reach a good agreement between the two estimates of $\langle A^2
\rangle$, the $\Lams$ parameter being taken from previous works to be the  same
for both two- and tree-point Green function matchings. Our estimate, $\beta_2 =
7400(1500)$ was about twice larger than  the result $\beta_2 = 3088$ in
\cite{Chety2}.  Still this fit went  in the right  direction, whence the authors
of \cite{Chety2} expected their result  to lead to a fair fit to lattice data.  
 
This expectation turns out to be correct :

  {\it First, the ratio of the two 
estimates of $\sqrt{\langle A^2 \rangle}$ is equal to $1.21(18)$, i.e. compatible
with 1,  provided the leading Wilson coefficients are consistently expanded at
the three-loop level and the subleading  coefficients of $\langle A^2 \rangle$
are computed to the leading logarithms.  Second, in the same {\bf joint} fit,
$\Lams$ is
 estimated to be $233(28)MeV$}, in amazing  agreement with previous  estimates of
$\Lams$ appearing in the literature (see for instance \cite{alpha,propag}). Thus, the
present analysis ends up with a twofold success and we can conclude that  OPE  leads to  a good description of
the deviations of the running coupling constant and of the gluon propagator  from
their perturbative behaviour in terms of perturbatively available coefficients 
multiplying one phenomelogical condensate: the sole non vanishing 
non-perturbative contribution up to the order $1/p^2$, namely the gluon
condensate  $\langle A^2 \rangle$.

However, for this  OPE description  to be consistent, it is unambiguously
demonstrated that the leading coefficient must be taken to three loops; on the
contrary there is  a clear failure at two loops, even
though our analysis has been performed at an unusually large energy scale, up to
10 GeV. As was discussed in the
preliminary study,  such a disagreement indicates that we are, in this case,  in
the situation described in \cite{MartiSach}, i.e. that the perturbative order is
too low to give an acceptable precision in the estimate of the power
corrections.  Taking into
account power corrections does not  make sense unless the leading contribution is
computed perturbativeley to a sufficient  accuracy. In a simple mathematical
language, it makes no sense to  consider the $1/p^2$  corrections when one does
not consider the $1/\ln(p^2)$ corrections to a sufficient order. OPE is often used
with the leading coefficients only known to two loops (sometimes to one loop).
We believe in view of our results that these attempts should be reconsidered
with care.

As for us, we were in a
particularly favourable situation to   analyse the problem thoroughly.  
We have rather accurate results. The dimension $2$ power correction clearly
shows up clearly and can be fully consistently attributed to an  $A_\mu A^\mu$ 
condensate in the Landau gauge in full agreement with the theoretical
expectations . Since we are working in the Landau gauge we produce (and use  later on) 
{\bf bare} gauge field configurations which minimize $A^\mu A _\mu $ with respect to the gauge group. 
This has the interesting consequence that the quantities we measure
 are invariant under infinitesimal gauge transformations in the vicinity of the
 Landau gauge.
Still the link between what we call the $\langle A^2\rangle$  and the  $\langle A^2_{\rm min}
\rangle$ defined in~\cite{Zakh} should be better clarified,
 which implies a better understanding of the renormalization procedure\footnote{it is a pleasure to acknowledge discussions with V.I Zakharov on this topic}. Taking such a direct link for granted, we can estimate from Eqs. (\ref{threeloops}) the tachyonic gluon  mass defined in ref.
\cite{Zakh},   to  be $\sim .8$ GeV\footnote{We recall that all scale-dependent quantities are evaluated at $\mu=10$~GeV.}.
Using instead the notion of critical mass,  $M^2_{\rm crit}$, introduced in ref.
\cite{Novi}, which is the scale at which the non-perturbative condensate
contributes $10\%$ of the total, we estimate it for the gluon propagator to be
$\sim 2.6$ GeV. Both these scales express a rather large contribution from the
$A^2$  condensate. %Do they point to the existence in QCD of some
%``intermediate'' scale between  $k^2$ and $\Lambda_{\rm QCD}^2$ \cite{Zakh} ? We
%are unable to say much about that. 

\section{Acknowledgments} 
We specially thank D. Becirevic for thorough discussions at the earlier stages of the 
work. We are grateful to V.I. Zakharov for his illuminating remarks. We are also indebted to 
Y. Dokshitzer and G. Korchemsky 
for several inspiring comments and to F. Di Renzo for clarifying the present status of their
work  concerning power corrections in Wilson loops. J. R-Q thanks J. 
 Rodr\'{\i}guez Garc\'{\i}a and is indebted to Spanish 
Fundaci\'on Ram\'on Areces for financial support.
These calculations were performed on the QUADRICS QH1 located 
in the Centre de Ressources
 Informatiques (Paris-sud, Orsay) and purchased thanks to a
  funding from the Minist\`ere de
  l'Education Nationale and the CNRS.

\appendix

\section{One-loop anomalous dimension of $A^2$}

\begin{figure}[hbt]
%\vspace*{-1.cm}
\hspace*{-1.3cm}
\begin{center}
%\begin{tabular}{ll}
%\epsfxsize6.5cm\epsffile{propag.eps} & 
\epsfxsize9.5cm\epsffile{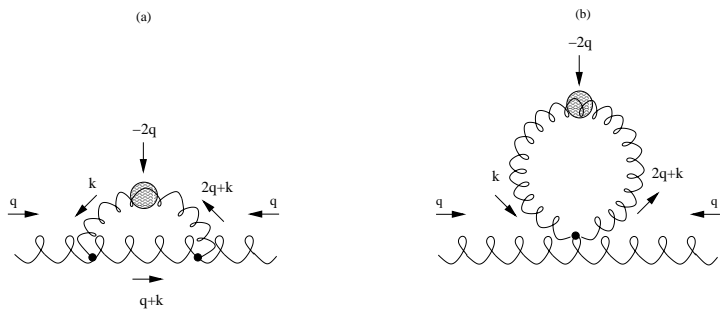} % \\
%\end{tabular}
\caption{\small {\it The graphs involved in the  computation of the anomalous dimension of $A^2$. The cross-hatched blobs 
 indicate the insertion of the $A^2$ operator, the dots are ordinary QCD vertices.}}
\label{Fig4}
\end{center}
\end{figure}

\hspace*{\parindent}
The task of computing the one-loop anomalous dimension of the matrix element 
$\langle A^2 \rangle$ (that of the local operator itself is directly obtained from 
the former as explained in section \ref{sec:model}~)  requires only to isolate the 
 UV divergent part of the diagrams in Fig. \ref{Fig4}. We follow dimensional 
regularization prescriptions to write:

\beq
\left[\Gamma_a\right]^{\mu \nu}_{a b}(q,-q)=\left(g^{\mu \nu}-\frac{q^\mu q^\nu}{q^2} \right)
\delta_{a b} \left( \frac{1}{\varepsilon} \left\{ 3 N_c \frac{\alpha_b}{4 \pi} \ + 
O\left(\alpha_b^2 \right)  \right\} \ + \dots \right) \; \; , 
\nonumber \\
\left[\Gamma_b\right]^{\mu \nu}_{a b}(q,-q)=\left(g^{\mu \nu}-\frac{q^\mu q^\nu}{q^2} \right)
\delta_{a b} \left( \frac{1}{\varepsilon} \left\{ -\frac92 N_C 
\frac{\alpha_b}{4 \pi} \ + O\left(\alpha_b^2 \right) \right\} \ + \dots \right) \; ;  
\label{UVpoles}
\eeq

\noindent where an a(b)-like diagram in Fig. \ref{Fig2} with amputated 
external gluon legs is denoted by $\Gamma_{a(b)}$ and where $\epsilon\equiv 2-d/2$.
The particular kinematics  we choose (see the figure), where the incoming
momentum flow is non vanishing,   eliminates automatically the IR divergences
and makes the UV analysis easier. We require  the final result for amputated
diagrams to be transverse to the external momenta, but  this is merely a
convention to be also applied to the tree-level term. Furthermore, had we
considered two different incoming momenta, the tensors in  Eq. (\ref{UVpoles})
would have acquired a more complicated form. 

%%%%%%%%%%%%%%%%%%%%%%%%%%%%%%%%%%%%%%%%%%%%%%%%%%%%%%%%%%%%%%%%%%%%%%
%%%%%%Figure 4
%%%%%%%%%%%%%%%%%%%%%%%%%%%%%%%%%%%%%%%%%%%%%%%%%%%%%%%%%%%%%%%%%%%%%
%%%%%%%%%%%%%%%%%%%%%%%%%%%%%%%%%%%%%%%%%%%%%%%%%%%%%%%%%%%%%%%%%%%%%%%

This kind of IR regularization, by imposing a non-null incoming momentum flow to the local 
operator, leads of course to UV poles results equivalent to those obtained from any other one.
We have tested this by considering a null incoming momentum flow and both introducing a 
certain cut-off to regularize IR divergences and separating IR and UV poles obtained by 
dimensional regularization.

If we collect the  tree-level results and those from Eqs. (\ref{UVpoles}), we can write:

\beq
\langle g^a_\mu | A^2 | g^b_\nu \rangle \ = \ 2 
\left( g_{\mu \nu} - \frac{q_\mu q_\nu}{q^2} \right) \delta^{a b}
\left( 1 + \frac{1}{\varepsilon} 
\left\{ \frac{3 N_C}{4} \frac{\alpha_b}{4 \pi} \ + \ O\left(\alpha_b^2 \right) \right\}
\ + \dots \right)
\label{UVpoleFin}
\eeq

\noindent where the matrix element in l.h.s. of Eq. (\ref{UVpoleFin}) is defined
for  explicitly cut external gluons. Combinatorics gives a multiplicity
factor $2$ for a-like  diagram, $1$ for b-like, which have been taken into
account in the last result. 

We should now renormalize the matrix element 
defined in Eq. (\ref{UVpoleFin}). Our aim being to determine its anomalous dimension computation to only
one-loop, the simple MS prescription of simply dropping away from bare quantities  the 
poles for $\varepsilon \to 0$ can be applied. Discrepancies between such a prescription and 
MOM or any other one  appear only beyond one-loop. Then we will have :

\beq
\widehat{Z}^{MS} & = & 1 \ + \ \frac{1}{\varepsilon} 
\left( \frac{3 N_C}{4} \ \frac{\alpha_b}{4 \pi} 
+ O\left( \alpha_b^2 \right) \right) \nonumber \\ 
& = & 1 \ + \ \frac{1}{\varepsilon} 
\left( \frac{3 N_C}{4} \ \frac{\alpha^{MS}(\mu)}{4 \pi} 
+ O\left( \alpha^{MS}(\mu) \right)^2 \right) \ .
\label{ZMS}
\eeq

\noindent From Eq. (\ref{ZMS}), the anomalous dimension can be written as 
(see for instance \cite{Davy}):

\beq
\widehat{\gamma}^{\rm MS}\left(\alpha^{MS}(\mu)\right) \ = \ 
\frac{d}{d\ln{\mu^2}}\ln{\widehat{Z}^{\rm MS}}
\ \nonumber \\ 
= - \frac{3 N_C}{4} \ \frac{\alpha^{MS}(\mu)}{4 \pi} + 
O\left( \left(\alpha^{MS}(\mu)\right)^2 \right) \ \ . 
\label{gMS}
\eeq

Thus  we  obtain in the MOM scheme,

\beq
\widehat{\gamma}\left(\alpha(\mu)\right) \ = \
- \frac{3 N_C}{4} \ \frac{\alpha(\mu)}{4 \pi} + 
O\left( \left(\alpha(\mu)\right)^2 \right) \ \ . 
\label{gMOM}
\eeq
From this  we deduce finally, including the gluon anomalous dimension
\beq
\gamma_{A^2}\left(\alpha(\mu)\right) \ = \
- \frac{35 N_C}{12} \ \frac{\alpha(\mu)}{4 \pi} + 
O\left( \left(\alpha(\mu)\right)^2 \right) \ \ . 
\label{ga2}
\eeq

\end{document}